\documentclass[conference]{IEEEtran}
\IEEEoverridecommandlockouts
\usepackage{cite}
\usepackage{amsmath,amssymb,amsfonts}
\usepackage{algorithmic}
\usepackage{graphicx}
\usepackage{textcomp}
\usepackage{xcolor,comment}
\usepackage{multicol, multirow}
\usepackage{float}
\usepackage{array}
\newcolumntype{L}{>{\centering\arraybackslash}m{2.0cm}}
\def\BibTeX{{\rm B\kern-.05em{\sc i\kern-.025em b}\kern-.08em
    T\kern-.1667em\lower.7ex\hbox{E}\kern-.125emX}}

\renewcommand{\baselinestretch}{0.99}

\newcommand{\Xn}{{{\cal{X}}_n}}
\newcommand{\Y}{{\{y_s\}_{s=1}^S}}
\newtheorem{theorem}{Theorem}
\newtheorem{exam}{Example}

\begin{document}

\title{Maximum Likelihood Quantum Error Mitigation 
for Algorithms with a Single Correct Output }
\author{\IEEEauthorblockN{Dror Baron}
\IEEEauthorblockA{\textit{NC State University}\\
Raleigh, NC, USA \\
dzbaron@ncsu.edu\\}
\and
\IEEEauthorblockN{
Hrushikesh Pramod Patil}
\IEEEauthorblockA{\textit{NC State University}\\
Raleigh, NC, USA \\
haptil2@ncsu.edu\\}

\and
\IEEEauthorblockN{
Huiyang Zhou}
\IEEEauthorblockA{\textit{NC State University}\\
Raleigh, NC, USA \\
hzhou@ncsu.edu\\}
}

\maketitle

\begin{abstract}
Quantum error mitigation is an important technique to reduce the impact of noise in quantum computers. 
With more and more qubits being supported on quantum computers, there are two emerging fundamental challenges. 
First, the number of shots required for quantum algorithms with large numbers of qubits needs to increase in order 
to obtain a meaningful distribution or expected value of an observable. Second, although steady progress has been made in
improving the fidelity of each qubit, circuits with a large number of qubits are likely to produce erroneous results. 
This low-shot, high-noise regime calls for highly scalable error mitigation techniques. In this paper, we propose a 
simple and effective mitigation scheme, qubit-wise majority vote, for quantum algorithms with a single correct output. 
We show that our scheme produces the maximum likelihood (ML) estimate under certain assumptions, 
and bound the number of shots required. Our experimental results on real quantum devices confirm that our proposed 
approach requires fewer shots than existing ones, 
and can sometimes recover the correct answers even when they are not observed from the measurement results.
\end{abstract}


\section{Introduction}

Quantum error mitigation has received a great deal of attention.
However, as we move to quantum circuits of hundreds of qubits and beyond in the 
noisy intermediate scale quantum (NISQ) era \cite{Kim2023}, 
we must pay close attention to the number of shots required and the computational complexity of the error mitigation method.

Popular quantum error mitigation approaches, including probabilistic noise cancellation (PEC) \cite{PhysRevLett.119.180509_Pec}, zero-noise extrapolation (ZNE) \cite{9259940_zne}, 
and Clifford data regression (CDR) \cite{Czarnik2021errormitigation_cdr}, have a high sampling overhead, 
and do not scale well as the number of qubits increases. 
In contrast, the M3 approach \cite{PRXQuantum.2.040326_m3} is computationally tractable. 
However, this efficiency relies on some assumptions. For example, 
M3 only considers outputting strings that were actually measured. 
Therefore, as the number of qubits, $n$, increases, the number of shots, $S$, that
the quantum algorithm performs must grow exponentially with $n$. Such $S$ are impractical beyond a few dozen qubits.

Our goal in this line of work is to study a low-complexity quantum error mitigation
scheme that can scale to a large number of qubits, $n$, while offering some
optimality properties. 

To simplify our analysis, we focus on quantum algorithms that 
in the hypothetical noise-free setting should have a single correct output. 
One prominent example of an algorithm with a single correct output is Bernstein-Vazirani (BV) \cite{doi:10.1137/S0097539796300921_bv}.
For such algorithms, we propose a simple {\em qubit-wise majority vote} (QMV), and prove its optimality under certain technical conditions. 
We further provide bounds on the number of required shots, $S$.
In addition to requiring fewer shots than some other schemes, QMV does not rely explicitly on calibration data 
from quantum devices.
To demonstrate the potential for moving beyond quantum algorithms that have a single correct output, 
in Section \ref{subsec:multiple} we propose a scheme that extends QMV in order to handle two antipodal outputs.

Our experimental results on real quantum devices confirm that our proposed 
approach requires fewer shots than existing ones, 
and can sometimes recover the correct answers even when they are not observed from the measurement results.

The rest of the paper is organized as follows. Section~\ref{sec:prior_art} describes the problem formulation
and provides background content.
Section \ref{sec:main} presents our main ideas including the QMV approach, 
and we provide associated derivations in Section~\ref{sec:derivation}. 
Next, Section \ref{sec:exp_methodology} describes our experimental methodology and associated results.
Finally, we conclude in Section \ref{sec:conculsion}.

\section{Problem formulation and prior art}\label{sec:prior_art}

After describing our problem formulation in Section~\ref{sec:problem_formulation},
some related quantum error mitigation schemes
are overviewed in Section~\ref{subsec:q_error_mit}.
Our proposed approach has statistical properties that appear in the 
estimation literature, which are described in Section~\ref{subsec:estimation}.

\subsection{Problem formulation}\label{sec:problem_formulation}

Quantum algorithms operate on quantum states, which reside in a Hilbert space. For an $n$-qubit system, the Hilbert space 
has dimension $2^n$. The outcome is measured in an $n$- dimension binary space $\Xn$,
\[
\Xn = \{0,1\}^n.
\]

Our focus is on quantum algorithms that have a single correct output,
$x_0 \in \Xn$, where this output $x_0$ is unknown.
Unfortunately, the quantum system outputs noisy versions of $x_0$. Specifically,
we are given $S$ noisy shots,
\[
Y= \Y,
\]
where $y_s \in \Xn$.
We further assume that the probabilistic mechanism that governs the
dependence of $y_s$ on $x_0$ is known, and assume that the $S$ shots, $\Y$, are 
{\em independent and identically distributed} (i.i.d.).
Given conditional probabilities that govern the dependence of $y\in\Xn$ on $x\in \Xn$,
\begin{equation}
\label{eq:noise}
\Pr(y|x), \forall x, y \in \Xn,
\end{equation}
our goal is to estimate the unknown $x_0$.

\subsection{Quantum error mitigation}
\label{subsec:q_error_mit}

Matrix free measurement mitigation (M3)\cite{PRXQuantum.2.040326_m3}
focuses on the need for fast error mitigation algorithms when the number of qubits
is substantial. The authors consider a hypothetical matrix, $A$, that relates probabilities of different quantum outputs
in the ideal case without noise, and the noisy case.
In principle, all the probabilities that correspond to all entries of $A$ could be positive.
However, because the number of shots, $S$, is finite, only part of the (noiseless, noisy) pair 
events are encountered among the shots, which lets M3 use a smaller $\widetilde{A}$ matrix.
As the M3 matrix-free measurement mitigation name suggests, there is no need to store $\widetilde{A}$ explicitly.
Instead, the noise-free probability vector is obtained by using $\widetilde{A}$ implicitly. 

The authors put special emphasis on somewhat large problems. 
However, in truly large problems with many dozens of qubits,
some outputs will never occur among the $S$ shots.
For quantum algorithms that (without noise) can only produce a small number of possible outputs,
some of these correct outputs may never be measured.
Indeed, as $n$ increases, it becomes more likely that the true outputs will not be measured.
But if none of the correct outputs are measured, then the $\widetilde{A}$ matrix simply does not include 
information about these correct values, and they will be ignored by the algorithm. 

The shortcomings of M3 are shared by algorithms such as Qbeep \cite{10.1145/3579371.3589043_qbeep} and 
HAMMER \cite{10.1145/3503222.3507703_hammer}. 
HAMMER performs correction on the observed probability distribution by assigning weights based on the Hamming distance of the observed bitstrings. 
QBeep resembles HAMMER, but uses a Poisson distribution and device calibration data to calculate the weights for the correction of the input distribution.

M3, HAMMER and QBeep process the output probability distributions. In contrast, zero noise extrapolation (ZNE) \cite{9259940_zne},
Clifford data regression (CDR) \cite{Czarnik2021errormitigation_cdr}, and probabilistic error cancellation (PEC) \cite{PhysRevLett.119.180509_Pec} 
predominantly mitigate the errors in the observables. However, the aforementioned approaches  
lack scalability with respect to the sampling overhead, which scales exponentially with the number of qubits, $n$.

While M3 can approach larger $n$ than some other error mitigation approaches, 
because it will not be able to produce the correct output if it is never actually measured, 
its sweet spot is still a somewhat moderate $n$.
This moderate $n$ is large enough to be computationally challenging
for some other algorithms, yet small enough to ensure that the correct outputs are usually measured.
Finally, because M3 can process larger $n$ than some other error mitigation procedures, we compare our approach
(described in Section III) mainly to M3. We will show that our approach offers Bayes-style error mitigation
quality that relies implicitly on a probabilistic model, while also addressing the large $n$ case.

\subsection{Estimation}
\label{subsec:estimation}

Our approach borrows heavily from the literature on signal and parameter estimation.
Therefore, we provide background on possible estimators for the unknown input $x$.

{\bf Mode.}
A simple insight in quantum error mitigation \cite{cai2023quantum_err_mit}
is that $y$ will be more likely when it is similar or close to $x_0$ in some manner.
For example, $y$ that have small Hamming distance from $x_0$ might be more likely.
Because nearby $y$ are more likely, if we collect many shots, $S$,
then it is likely that most of $Y$ will be vectors somewhat close to $x_0$.
Specifically, many vectors close to $x_0$ will be measured multiple times.

Among all the vectors $y \in \Xn$ and under  
some technical conditions for the noise mechanism, $\Pr(y|x)$ (\ref{eq:noise}),
the most likely vector to be measured is $x_0$ itself. Therefore, if $S$ is large enough, 
then it is plausible that we will measure $x_0$ more than any other vector.

The mode estimator is based on this idea. 
We define the number of times that some vector
$x \in \Xn$ was measured,
\[
f_M(x) = \sum_{i=1}^N 1_{\{y_i=x\}},
\]
where $1_{\{\cdot\}}$ is an indicator function, and the subscript $M$ is for the mode.
The mode estimator is now defined,
\begin{equation}
\label{eq:xhat_mode}
\widehat{x}_M = \arg\max_{x \in \Xn} f_M(x).
\end{equation}

Unfortunately, in many applications
$|\Xn| = 2^n \gg S$, and there may not be any $x$ that is measured multiple times, i.e., $f_M(x)>1$. 
Even if a few $x$ are measured multiple times, they need not be the true $x_0$.
Therefore, in many applications, the mode estimator requires an excessively large $S$ to succeed.

{\bf Maximum likelihood.}
A more precise approach computes the probabilities of the $M$ measurements conditioned on a hypothetical
$x \in \Xn$, and then selects the estimator $\widehat{x}$ that maximizes the conditional probability.
This approach is known as the maximum likelihood (ML) estimator. 

Formally, we define $f_{ML}(x)$ to be the likelihood or probability of $Y$ conditioned on $x$,
\begin{equation}
\label{eq:ML}
f_{ML}(x) 
= \Pr(Y|x)
= \Pr(\{y_i\}_{s=1}^S|x)
= \prod_{s=1}^S \Pr(y_s|x),
\end{equation}
where we are implicitly relying on the assumption that measurements obtained from
different shots are independent.
As before, the ML estimator selects the $x$ with the largest likelihood,
\begin{equation}
\label{eq:xhat_ML}
\widehat{x}_{ML} = \arg\max_{x \in \Xn} f_{ML}(x).
\end{equation}

{\bf Maximum a posteriori.}
What if we believe that some $x$ are more likely than others?
This could happen if we have run our quantum algorithm many times in the past and seen that some outputs are more likely;
or we might have some side information about our quantum system.
To see why some outputs being more likely a priori than others may create complications, 
consider the following example.

\begin{exam}
Suppose that $f_{ML}(x_a)$ is 1.01 times larger than $f_{ML}(x_b)$,
but a priori $x_b$ is $10^6$ more likely than $x_a$.
Based on the measurements in $Y$, $x_a$ is only mildly more likely, yet when also incorporating our 
prior information, $x_b$ becomes much more likely. $\square$
\end{exam}

To quantify these notions precisely, suppose that we have prior probabilities for $x \in \Xn$, 
denoted by $\Pr(x)$, and consider the probability of $x$ conditioned on the data $Y$,
\begin{eqnarray*}
\Pr(x|Y)
&=&
\frac{\Pr(x,Y)}{\Pr(Y)} \\
&=&
\frac{\Pr(x,Y)}{\sum_{x' \in \Xn} \Pr(x',Y)} \\
&=&
\frac{\Pr(x)\Pr(Y|x)}{\sum_{x' \in \Xn} \Pr(x')\Pr(Y|x')} \\
&=&
\frac{\Pr(x) f_{ML}(x)}{\sum_{x' \in \Xn} \Pr(x')  f_{ML}(x')}. \\
\end{eqnarray*}
Maximizing this conditional probability, $\Pr(x|y)$, yields the {\em maximum a posteriori} (MAP)
estimator. Formally, we define
\[
f_{MAP}(x) = \Pr(x|Y)
\]
and
\begin{equation}
\label{eq:xhat_MAP}
\widehat{x}_{MAP} = \arg\max_{x \in \Xn} f_{MAP}(x).
\end{equation}
Note that if all $x \in \Xn$ have the same prior, 
i.e., $\Pr(x)=2^{-n},\ \forall x \in \Xn$,
then the MAP and ML estimators coincide.

\section{Maximum likelihood mitigation}
\label{sec:main}

This section provides our main ideas. Throughout, we consider i.i.d. measurement noise.
In Section~\ref{sec:QMV}, bit flips are symmetric, 
i.e., $\Pr(y=1|z=0)=\Pr(y=0|z=1)=p$, and
the quantum algorithm has a single correct output, $x_0 \in \Xn$. 
For this symmetric i.i.d. model, Theorem~\ref{th:main}
will show that the optimal ML estimator is obtained by performing a qubit-wise majority vote (QMV)
among the $S$ shots for each of the $n$ qubits. QMV
is a simple estimator, yet it is ML-optimal for symmetric i.i.d. noise.
The number of shots required by QMV
is described in Section~\ref{subsec:num_shots}. QMV
requires $S=O(\ln(n)/(0.5-p)^2)$, which is logarithmic in $n$. 
In contrast, M3 requires
$S=O((1-p)^{-n})$, which is exponential.

Next, Section~\ref{sec:assymetric} considers asymmetric noise, i.e.,
$\Pr(y=1|z=0)=p_{01}$ and $ \Pr(y=0|z=1)=p_{10}$.
We show that a qubit-wise vote is still ML-optimal, but we may need to move away from a
simple majority decision based on how much $p_{01}$ and $p_{10}$ differ.

Section~\ref{subsec:multiple} shows how to extend QMV and apply it to quantum algorithms with two possible antipodal outputs.
While a result for two antipodal outputs is narrow in scope, it shows that there is hope for quantum algorithms with
multiple correct outputs.
Finally, Section~\ref{sec:subsetting} discusses the case where QMV yields 
a close result on some qubits, and adaptive measurement subsetting extends 
Das et al. \cite{10.1145/3466752.3480044_MSubset} by allocating more measurements 
to the qubits with close votes.

\subsection{Qubit-wise majority vote (QMV)}
\label{sec:QMV}

The quantum algorithm that operates on the Hilbert
space spanned by $\Xn$ is assumed to have a single correct output, $x_0 \in \Xn$. 
This output $x$ is unknown.
To compute the ML estimate, and using (\ref{eq:ML}) and (\ref{eq:xhat_ML}),
\begin{equation}
\label{eq:ML_revisited}
\widehat{x}_{ML} = \arg\max_{x \in \Xn} f_{ML}(x)
= \arg\max_{x \in \Xn} \prod_{s=1}^S \Pr(y_s|x).
\end{equation}
Because we assume that measurements obtained from different qubits
are independent, we have $\Pr(y_s|x) = \prod_{i=1}^n \Pr(y_{si}|x_i)$,
and so
\[
f_{ML}(x) = \prod_{s=1}^S \prod_{i=1}^n \Pr(y_{si}|x_i).
\]
This formulation relies on the i.i.d. assumption. However, real-world quantum circuits
may feature qubits with varying circuit depths, and shallow qubits could be less noisy.
More accurate modeling of noise and the resulting error mitigation schemes are beyond the scope of our study.

One way to identify $f_{ML}(x)$ is to consider all possible strings, 
and find the one that maximizes the likelihood. However, considering all possible strings requires
exponential complexity. 
Instead, we prove that QMV is the ML estimate.
The proof appears in Section~\ref{subsec:th:main:proof}.

\begin{theorem}
\label{th:main}
Consider a quantum algorithm that has a single correct output, $x_0\in\Xn$,
where the measurements obtained by all shots are statistically independent,
and the probabilities of flipping any entry, $i\in\{1,\ldots,n\}$, are symmetric
and less than $0.5$, $\Pr(y_i=1|z_i=0)=\Pr(y_i=0|z_i=1)=p<0.5$.
Then $\widehat{x}_{ML}$ for entry~$i$ can be obtained
by performing QMV on $\{y_{si}\}_{s=1}^S$.
\end{theorem}

{\bf Discussion.}
Let us make several comments about our theorem.
First, what is QMV on $\{y_{si}\}_{s=1}^S$?
Considering entry $i$ (i.e., qubit $i$) in each of the $S$ vectors measured by the $S$ shots,
we have $S$ bits. To keep the presentation simple, suppose that these bits have values
0 and 1 (different quantum measurements might be feasible in some systems, e.g., $\pm 1$),
and denote the number of zeros by $N_{0i}$, hence the number of ones, which we denote by $N_{1i}$,
satisfies $N_{0i}+N_{1i}=S$. If $N_{0i}>N_{1i}$ then we declare the outcome to be 0, else 
we declare it to be 1. As QMV is performed on every measured qubit, the complexity is $O(nS)$.

Second, why does QMV work? 
We provide an example that may provide some insight.

\begin{exam} 
\label{ex:low_prob_error}
Consider an example where $n=5$, $x_0=00000$, 
the probability that a qubit is flipped is $0.2$, and $S=10$. 
Because $\{y_{si}\}_{s=1}^S$ is comprised
of $S=10$ measured outcomes, we can evaluate the probability that at least 5 of the 10 are flipped,
\begin{eqnarray*}
\Pr(\text{error})
&=&
\sum_{f=5}^{10}\Pr(\text{$f$ flips}) \\
&=&
\sum_{f=5}^{10} 0.2^f 0.8^{10-f} \text{Choose}(10,f) \\
&=&
0.0328.\end{eqnarray*}
Therefore, QMV will estimate the qubit
correctly with a probably of $96.72\%$. $\square$
\end{exam}

Third, why does our theorem require that bit flips have probability 
less than $0.5$?
If any entry~$i$ has bit flip probability that exceeds 0.5, then
QMV on $\{y_{si}\}_{s=1}^S$ is likely to provide the wrong answer.
Indeed, if the probability of bits being flipped is somewhat close to 0.5, 
then we may need to increase $S$.
However, unless this probability is quite close to 0.5, 
$S$ need not be too large (Section~\ref{subsec:num_shots_derivation}).
The requirement that for each entry~$i$ the probability
that entry~$i$ is flipped is less than $0.5$ seems 
plausible in many quantum settings, and our theorem applies to quantum algorithms 
such as Bernstein-Vazirani (BV) \cite{doi:10.1137/S0097539796300921_bv}
that have a single correct output.

Fourth, why does the theorem require a quantum algorithm that has a single correct output?
To see why, let us consider a modified version of Example~\ref{ex:low_prob_error}.

\begin{exam}
We now have two possible outputs,
$x_a=00000$ and  $x_b=11111$,
and they each occur with probability 0.5. 
The distribution of the number of ones in
$\{y_{si}\}_{s=1}^S$ will be centered around $S/2$.
(An algorithm that processes this specific example appears in Section~\ref{subsec:multiple}.)
$\square$
\end{exam}

Finally, consider a scenario where all $2^n$ possible values for $x$ are equi-probable.
In this case, the maximum a posteriori 
(MAP) estimator is identical to the ML estimator. As a corollary, when values of $x$ are
equi-probable, the MAP estimator for entry~$i$ can be obtained
by performing QMV on $\{y_{si}\}_{s=1}^S$.
On the other hand, if there is a single correct output, $x$, 
but the $2^n$ different possible outputs have different priors, 
then MAP estimation must utilize information about these priors in the decision process. 
This point that MAP must consider the prior is illustrated in the following example.

\begin{exam}
Suppose that there is a prior knowledge that entry~$1$ in the output of the quantum
algorithm is {\em always} $0$ (i.e., if we measure $1$, then it is an error).
It could be that $y_{s1}=1$ occurs more frequently than $y_{s1}=0$, but we know that these are errors and can discard them. This is the approach employed in quantum assertions \cite{DBLP:conf/asplos/LiuBZ20}.
$\square$
\end{exam}

\begin{figure*}
    \centering
    \includegraphics[width=\linewidth]{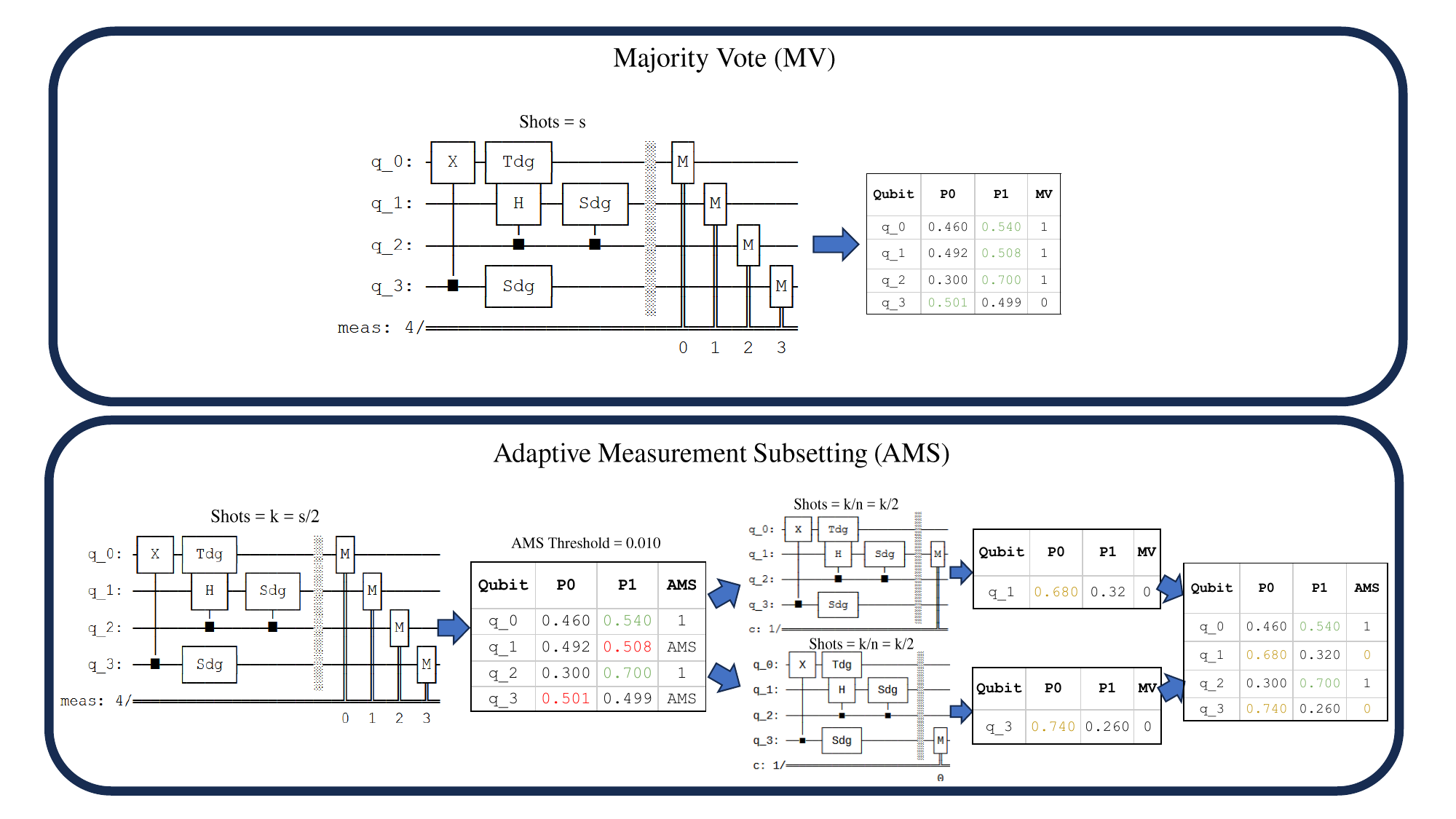}
    \caption{Example for the procedure of the proposed qubit-wise majority vote (QMV) and adaptive measurement subsetting (AMS)
    approaches on an example 4 qubit random circuit. For AMS the same example is repeated and the threshold is selected to be 0.01.}
    \label{fig:MLEandAMS}
\end{figure*}

\subsection{Reduction in number of shots}
\label{subsec:num_shots}

How many shots does QMV need?
For each bit, there are two types of error.
The first type involves the ground truth being 0,
while QMV votes 1.
In Section~\ref{subsec:num_shots_derivation}, we will bound this 
probability expression, 
\begin{equation}
\label{eq:prob_type1}
\Pr(\text{QMV $1$}|\text{ground truth $0$}, S, p).
\end{equation}
The second type of error involves the ground truth being 1,
while QMV votes 0. Because 
$\Pr(y=1|z=0)=\Pr(y=0|z=1)=p$,
the probability for this second type of error equals 
the first, (\ref{eq:prob_type1}).

Similar to Theorem~\ref{th:main}, our derivation for the number of shots 
(Section~\ref{subsec:num_shots_derivation}) assumes that $p<0.5$.
It also assumes that $S$ is even; allowing for odd valued $S$ would
only require minor yet tedious changes to the derivation.
The end result is that using $S=\frac{0.5\ln(n)}{\epsilon^2}$
shots provides a probability $O((\ln(n))^{-0.5})$ that any of the $n$
qubits will vote erroneously, where  $\epsilon=0.5-p$. In other words, 
the closer we are to a 50--50 vote, the more shots we need.

While the number of shots required for QMV is logarithmic
in $n$, M3 requires an exponential number of shots. To see why, recall from
Section~\ref{subsec:q_error_mit} that M3 
will not be able to produce the correct output if it is never actually measured.
The correct output is measured with probability $(1-p)^n$,
hence the number of shots required is 
$S=O((1-p)^{-n})$. Therefore, we expect QMV to require much
fewer shots than M3 for large $n$.

\subsection{Asymmetric noise}
\label{sec:assymetric}

Theorem~\ref{th:main} proved that the ML estimator has a simple 
QMV form when the measurement noise is i.i.d. and
bit flips are symmetric.
What about asymmetric noise,
$\Pr(y=1|z=0)=p_{01}$ and $ \Pr(y=0|z=1)=p_{10}$?
Decoherence noise is asymmetric, as an excited state, $\left|1\right\rangle$, 
decays into a ground state, $\left|0\right\rangle$,
with a probability that depends on the latency and the relaxation time, $T_1$, 
of the device~\cite{nielsen2010quantum}.
Below, we will process asymmetric noise by modifying QMV
into a weighted vote, which is ML optimal. Before deriving the weighted vote, we provide an example.

\begin{exam}
To see why QMV is sub-optimal for asymmetric noise,
consider a scenario where $p_{01}=0.5$ and $p_{10}=0.0$.
That is, if the ground truth is 1, then we always measure 1.
However, if the ground truth is 0, then our measurements are random.
For a qubit with these bit flip probabilities, any result besides all-$1$ indicates that the ground
truth was 0, and QMV would often yield incorrect decisions.
$\square$
\end{exam}

We now provide our derivation for the asymmetric case, which extends the derivation used 
to prove Theorem~\ref{th:main} (Section~\ref{subsec:th:main:proof}).
Given asymmetric bit flip probabilities, $\Pr(y=1|z=0)=p_{01}$ and $ \Pr(y=0|z=1)=p_{10}$, 
and that among the $S$ measurements of a qubit $0$ is measured $a$ times and 
$1$ is measured $b$ ($= S-a$) times, we have posterior probabilities,
\[
f_{ML}(z_1=0)=\prod_{s=1}^S \Pr(y_{s1}|z_1=0) = p_{01}^b(1-p_{01})^a,
\]
\[
f_{ML}(z_1=1)=\prod_{s=1}^S \Pr(y_{s1}|z_1=1) = p_{10}^a(1-p_{10})^b.
\]
Therefore, the likelihood ratio,
$f_{ML}(z_1=1)$ over $f_{ML}(z_1=0)$, becomes 
\[
\frac{f_{ML}(z_1=1)}{f_{ML}(z_1=0)}
=\left( \frac{p_{10}}{1-p_{01}} \right)^a
\left( \frac{1-p_{10}}{p_{01}}\right)^b.
\]
When the ratio is larger than $1$, we take $1$ as the mitigated result,
else it is smaller than $1$, and we take $0$.
To reduce the computational complexity, consider the log likelihood,
\[
a\ln[(p_{10})/(1-p_{01})]-b\ln[(p_{01})/(1-p_{10})].
\] 
If the log likelihood is larger than $0$, we take $1$ as the mitigated result,
else it is smaller than $1$, and we take $0$.
In words, the asymmetric case involves a weighted vote. 
When $p_{01}=p_{10}$, i.e., symmetric bit flips, we have (unweighted) QMV.

\subsection{Adaptive measurement subsetting}
\label{sec:subsetting}

If QMV yields a close result on some qubits, we propose a technique called 
{\em adaptive measurement subsetting} (AMS), which extends the measurement subsetting approach 
by Das et al. \cite{10.1145/3466752.3480044_MSubset}. 
The key idea is to allocate more shots to measure qubits with close votes. In words, 
AMS maintains the circuit unchanged, but selectively measures subsets of qubits that exhibited close votes.
Each subset is comprised of a few of these qubits, and we measure one subset at a time. 
In addition to reducing the measurement crosstalk, measuring a small subset of qubits prompts the compiler to map 
the circuit so that these measured qubits are mapped to the physical qubits with the lowest measurement error rates. 

Instead of increasing the number of shots, we apply subsetting by first allocating half the 
shots to measuring all the qubits. The remaining half is then allocated to measuring subsets of qubits, one or a few qubits at a time. 
This subsetting approach is contrasted with QMV in Fig \ref{fig:MLEandAMS}. 
In the figure, both approaches have the same budget of $S$ shots. For AMS, half the shots, $k = S/2$, 
are allocated to the circuit where all the qubits are measured. For each qubit where 
$|p_0 - p_1|$ is below some threshold, 
where $p_0$ and $p_1$ are the empirical probabilities that the qubit was measured as 0 or 1, respectively, 
that qubit is selected for measurement subsetting. 
As the remaining shot budget is also $k = S/2$, and assuming that we have $m$ close vote qubits, each AMS circuit will have $k/m$ shots.

\subsection{Multiple correct outputs}
\label{subsec:multiple}

For quantum algorithms with multiple correct outputs, we must extend QMV. One extension is to use a sliding window to cover all the qubits and perform 
QMV within each window. For example, with a two-qubit sliding window, the majority vote is among $00$, $01$, $10$, and $11$. If we have prior knowledge 
that the quantum algorithm has two antipodal outputs (e.g., the GHZ state), or two symmetric solutions 
(e.g., bitstrings representing node partitions for a MAX-Cut problem), then the majority vote would be between $\{00, 11\}$ and $\{10, 01\}$. 
Using $n-1$ such two-qubit windows to cover $n$ qubits with one overlapping qubit between two windows, we can estimate the antipodal outputs.
In a numerical experiment, we use an $n=20$ qubit GHZ state with $S=4000$ shots and $p=35\%$ bit flips.
The all-zero and all-one states were rarely or never measured, hence the mode and M3 both fail to identify them. 
In contrast, the sliding window successfully identifies the all-zero and all-one states as the correct outputs. 
The detailed study of more general cases, including those with more than two correct outputs, is being considered in our ongoing work.

\section{Derivations}\label{sec:derivation}

\subsection{Proof of Theorem~\ref{th:main}}
\label{subsec:th:main:proof}

We derive an ML estimator to process the noisy measurements, $\Y$.
Recall from (\ref{eq:xhat_ML}) that
\[
\widehat{x}_{ML} = \arg\max_{z \in \Xn} f_{ML}(z)
= \arg\max_{z \in \Xn} \prod_{s=1}^S \Pr(y_s|z).
\]
Because $\Pr(y_s|z) = \prod_{i=1}^n \Pr(y_{si}|z_i)$, we have that 
\[
f_{ML}(z) = \prod_{s=1}^S \prod_{i=1}^n \Pr(y_{si}|z_i).
\]

Without loss of generality, suppose that we want to focus on an ML estimator for the first entry of $z$.
That is, $z$ at the other indices, $2$ through $n$, which we denote by $z_{\setminus 1}$, has been determined
separately, and we only need to decide between $z_1=0$ and $z_1=1$.
For $z_1=0$, the likelihood term is,
\begin{eqnarray*}
f_{ML}(z=0z_{\setminus 1})
\hspace*{-2.5mm}&=&\hspace*{-2.5mm}
\prod_{s=1}^S \prod_{i=1}^n \Pr(y_{si}|z_i) \\
\hspace*{-2.5mm}&=&\hspace*{-2.5mm}
\left[ \prod_{s=1}^S \Pr(y_{s1}|z_1=0) \right]
\hspace*{-1.5mm}
\left[ \prod_{s=1}^S \prod_{i=2}^n \Pr(y_{si}|z_i) \right],
\end{eqnarray*}
where $0z_{\setminus 1}$ denotes the concatenation of the bit $0$ (for $z_1$)
with $z_{\setminus 1}$. Note that the latter expression in square brackets
is a function of $z_{\setminus 1}$, which can be denoted by $f_{ML}(z_{\setminus 1})$,
\[
f_{ML}(z=0z_{\setminus 1})
=
\left[ \prod_{s=1}^S \Pr(y_{s1}|z_1=0) \right]
f_{ML}(z_{\setminus 1}).
\]
Similarly, for $z_1=1$, $f_{ML}(z=1z_{\setminus 1})$ can be expressed,
\[
f_{ML}(z=1z_{\setminus 1}) = 
\left[ \prod_{s=1}^S \Pr(y_{s1}|z_1=1) \right]
f_{ML}(z_{\setminus 1}).
\]
When maximizing the argument and having fixed $z_{\setminus 1}$,
the rest of $z$ at indices $2$ through $n$,
we need only perform a comparison between
$f_{ML}(z_1=0)=\prod_{s=1}^S \Pr(y_{s1}|z_1=0)$ and 
$f_{ML}(z_1=1)=\prod_{s=1}^S \Pr(y_{s1}|z_1=1)$.

Given that among the $S$ measurements of a qubit 
$0$ and $1$ appear $a$ and $b$ ($= S-a$) times, respectively,
\[
f_{ML}(z_1=0)=\prod_{s=1}^S \Pr(y_{s1}|z_1=0) = p^b(1-p)^a,
\]
\[
f_{ML}(z_1=1)=\prod_{s=1}^S \Pr(y_{s1}|z_1=1) = p^a(1-p)^b.
\]
Therefore, the ratio of $f_{ML}(z_1=1)$ over $f_{ML}(z_1=0)$ is
$(p)^{a-b}(1-p)^{b-a}$. 
For $p<0.5$, if $1$ appears more often than $0$ in the $S$ measurements of a qubit, 
then $b>a$, and the ratio is greater than $1$.
Similarly, if $0$ is the QMV result,
then $b<a$, and the ratio is less than $1$. 
$\square$

\subsection{Number of shots for QMV}
\label{subsec:num_shots_derivation}

We will bound 
$\Pr(\text{QMV $1$}|\text{ground truth $0$}, S, p)$
(\ref{eq:prob_type1}),
which is the probability of one type of error that
a qubit may experience,
and determine how many shots are needed in order for the
probability in any of the $n$ qubits to be small.
To simplify our presentation, we will use the notational 
convention, $\Pr(\text{QMV $1$}|0,S,p)$ for such expressions.
Note that
\begin{equation}
\label{eq:Pr_bound1}    
\Pr(\text{QMV $1$}|0, S, p)
= \sum_{a=0}^{c_1} \text{choose}(S,a) (1-p)^a p^{S-a},
\end{equation}
where $c_1=S/2$, because we are assuming that $S$ is even.
In words, $S/2$ or more of the votes are $1$,
and QMV incorrectly decides that the qubit was $1$,
despite the ground truth being $0$.

Next, denote
\[
f(a)= \text{choose}(S,a) (1-p)^a p^{S-a},
\]
for $a \in \{0, \ldots, S\}$.
Specifically, 
\begin{equation}
\label{eq:f(S/2)}
f(S/2) = \text{choose}(S,S/2) ((1-p)p)^{S/2}.
\end{equation}

Within our summation (\ref{eq:Pr_bound1}), 
the largest $\text{choose}(S,a)$ term is obtained for $a=c_1=S/2$.
Because $p<0.5$, the largest $(1-p)^a p^{S-a}$ term is also obtained for $S/2$.
The second largest term is for $S/2-1$,
\[
f(S/2-1) = \text{choose}(S,S/2-1) (1-p)^{S/2-1}p^{S/2+1}.
\]
Dividing these expressions,
\begin{eqnarray*}
\frac{f(S/2-1)}{f(S/2)} 
&=& 
\frac{ \text{choose}(S,S/2-1) } { \text{choose}(S,S/2)} 
\frac{(1-p)^{S/2-1}p^{S/2+1}} {((1-p)p)^{S/2}}\\
&=&
\frac{(S/2)!(S/2)!} {(S/2-1)!(S/2+1)!} \frac{p}{1-p} \\
&=&
\frac{S/2}{S/2+1} \frac{p}{1-p}.
\end{eqnarray*}
The ratio between $f(S/2-2)$ and $f(S/2-1)$,
\begin{eqnarray*}
\frac{f(S/2-2)}{f(S/2-1)} 
&=& 
\frac{ \text{choose}(S,S/2-2) } { \text{choose}(S,S/2-1)} 
\frac{(1-p)^{S/2-2}p^{S/2+2}}
{(1-p)^{S/2-1}p^{S/2+1}} \\
&=&
\frac{(S/2-1)!(S/2+1)!}{(S/2-2)!(S/2+2)!}
\frac{p}{1-p} \\
&=&
\frac{S/2-1}{S/2+2}
\frac{p}{1-p},
\end{eqnarray*}
is smaller, because
\[
\frac{S/2-1}{S/2+2} < \frac{S/2}{S/2+1}.
\]
Later ratios, such as between $f(S/2-3)$ and $f(S/2-2)$,
are even smaller.
Therefore, we can bound the summation for 
$\Pr(\text{QMV $1$}|0, S, p)$
in (\ref{eq:Pr_bound1}) as the largest term, $f(S/2)$, times a geometric sequence,
\[
1+b+b^2 + \ldots = \frac{1}{1-b},
\]
where 
\[
b = \frac{S}{S+2} \frac{p}{1-p} < \frac{p}{1-p}.
\]
Therefore, the geometric sequence sums to less than
\[
\frac{1}{1- \frac{p}{1-p}}
=\frac{1-p}{1-2p}.
\]
Therefore,
\begin{equation}
\label{eq:expression1}
\Pr(\text{QMV $1$}|0, S, p) <
f(S/2) 
\frac{1-p}{1-2p}.
\end{equation}

We can simplify this expression, (\ref{eq:expression1}), using
\begin{eqnarray}
\text{choose}(S/2) 
&=& 
\frac{S!}{(S/2)!(S/2)!} \nonumber \\
&=&
\frac{\left[\sqrt{2\pi S} (S/e)^S(1+\frac{1}{12S})\right]
[1+O(1/S^2)]}
{\left[\sqrt{2\pi (S/2)} ((S/2)/e)^{(S/2)}(1+\frac{1}{12(S/2)})\right]^2} \nonumber \\
&=&
\sqrt{\frac{2}{\pi S}} 2^S
(1-\frac{1}{4S})
[1+O(1/S^2)], \label{eqn:Stirling}
\end{eqnarray}
where Stirling's approximation for the factorial function was invoked.
Combining (\ref{eq:f(S/2)}), (\ref{eq:expression1}), and (\ref{eqn:Stirling}),
\begin{eqnarray}
\Pr(\text{QMV $1$}|0, S, p) 
\!&<&\!
\left[\!
\sqrt{\frac{2}{\pi S}} 2^S 
(1-\frac{1}{4S})\!\right] \nonumber \\
&\times&
((1-p)p)^{S/2} \nonumber \\
&\times&
\left[\frac{1-p}
{1-2p}\right] 
[1+O(1/S^2)]. \label{eq:expression2}
\end{eqnarray}
Because
\[
(1-1/(4S))[1+O(1/S^2)] < 1
\]
for sufficiently large $S$,
we have a simpler bound,
\[
\Pr(\text{QMV $1$}|0, S, p) 
<
(4(1-p)p)^{S/2} 
\sqrt{\frac{2}{\pi S}} 
\frac{1-p}{1-2p}.
\]

{\bf Discussion.}
The key term in our bound is
the exponential one,
$(4(1-p)p)^{S/2} $.
Denoting $p=0.5-\epsilon$,
\[
(4(1-p)p)^{S/2} 
=(4(0.5+\epsilon)(0.5-\epsilon))^{S/2}
=(1-4\epsilon^2)^{S/2}.
\]
The polynomial part is,
\[
\sqrt{\frac{2}{\pi S}} 
\frac{1-p}{1-2p}
=
\sqrt{\frac{2}{\pi S}} 
\frac{0.5+\epsilon}{2\epsilon}
=
\frac{1+\frac{1}{2\epsilon}}{\sqrt{2\pi S}}.
\]
Rewriting $S=\frac{c_2}{\epsilon^2}$ for some $c_2>0$,
the exponential and polynomial terms become
\[
(1-4\epsilon^2)^{\frac{c_2}{2\epsilon^2}} \approx e^{-2c_2}
\text{ and }
\frac{1+\frac{1}{2\epsilon}}{\sqrt{2\pi \frac{c_2}{\epsilon^2}}}
= \frac{\frac{1}{2}+\epsilon} {\sqrt{2\pi c}_2},
\]
respectively. Our entire expression is now
\[
\Pr(\text{QMV $1$}|0, S, p) 
<
\frac{\frac{1}{2}+\epsilon} {\sqrt{2\pi c_2}} e^{-2c_2}.
\]

Because we have $n$ qubits, 
we want $\Pr(\text{QMV $1$}|0, S, p)$ to be roughly $\frac{1}{n}$
(and ideally smaller than that). 
Using $c_2=0.5\ln(n)$,
\[
\Pr(\text{QMV $1$}|0, S, p) 
<
\frac{\frac{1}{2}+\epsilon} {\sqrt{\pi \ln(n)}} \frac{1}{n}.
\]
Different choices of $c_2$ can also be used.

While we have bounded 
$\Pr(\text{QMV $1$}|0,S,p)$,
$\Pr(\text{QMV $0$}|1,S,p)$ is identical, because 
we are considering symmetric noise,
$\Pr(y=1|z=0)=\Pr(y=0|z=1)=p$.
We conclude that 
$S=\frac{0.5\ln(n)}{\epsilon^2}$
is a good choice for the number of shots.

\section{Experimental results}\label{sec:exp_methodology}

\subsection{\bf Experimental setup}

We conduct our experiments on the IBM quantum devices. Each circuit in our experiments is optimized with the 
Qiskit \cite{Qiskit} transpiler level-3 optimization and run using the Qiskit runtime sampler primitive. We compare our algorithm 
with the state-of-the-art M3 error mitigation algorithm. The implementation of M3 was from the Python distribution of 
the M3 package. We also report the mode bitstring in the unmitigated probability distributions. 

{\bf Benchmarks.}
As our focus is quantum algorithms with single correct outputs, we use the Bernstein-Vazirani (BV) algorithm and random circuits (RC) in our evaluations. To guarantee a singular correct output for the random circuits, we initially set the circuit to a chosen bitstring (e.g., one with alternating 1s and 0s), then apply the random circuit 
followed by its inverse. We vary the number of qubits across all benchmarks and adjust the depth for the random circuit benchmark. We also vary the number of shots in each experiment to examine the impact of shots on circuits with high numbers of qubits. 

{\bf Evaluation criteria.}
As the circuits of interest have a single correct output, we use the Hamming distance to show how close the mitigated results are compared to the ideal noise-free solutions. A lower Hamming distance signifies a more favorable outcome.

\subsection{\bf Results}
Our results on the quantum device, \texttt{ibm\_sherbrooke}, are summarized in Table \ref{tab:BV} and Table \ref{tab:RandomCircuits}, for the BV and RC circuits, respectively. We can see that QMV consistently outperforms M3. Notably, it sometimes perfectly recovered
the correct solution, i.e., Hamming distance of 0, when the correct bit-string is not dominant or even unobserved in the noisy (i.e., unmitigated) results.

\begin{table}[!ht]
    \centering
    \caption{Comparison of Hamming distances for the dominant solutions in the Bernstein-Vazirani benchmark.}
    \begin{tabular}{|c|c|c|c|c|}
    
    \hline
        \textbf{Experiment Details} & \textbf{Shots} & \textbf{Mode}  & \textbf{M3} & \textbf{QMV} \\ \hline
        Qubits = 20 & 1024 & 6 & 6 & 6 \\ 
        Depth = 131 & 4096 & 7 & 7 & 5 \\ 
        2q Gates = 31 & 16384 & 2 & 2 & 3 \\ \hline
        Qubits = 30 & 2048 & 8 & 8 & 7 \\ 
        Depth = 185 & 8192 & 8 & 8 & 5 \\ 
        2q Gates = 46 & 32768 & 10 & 10 & 5 \\ \hline
        Qubits = 40 & 4048 & 15 & 11 & 8 \\ 
        Depth = 236 & 16384 & 10 & 10 & 6 \\ 
        2q Gates = 62 & 65536 & 10 & 10 & 5 \\ \hline
    \end{tabular}
    
    \label{tab:BV}
\end{table}

Table \ref{tab:BV} presents a comparison of the performance among the mode, M3, and QMV on the BV benchmark. The overall trend indicates a decrease in the Hamming distance across all methods as the number of shots increases. In general, the Hamming distance 
decreases more rapidly for QMV compared to both the M3 and the mode methods. 
The only exception where the mode and M3 reach a lower Hamming distance than QMV is because of a close vote on one qubit, leading QMV to produce an extra incorrect bit. 

\begin{table}[!htb]
    \centering
    \caption{Comparison of Hamming distance of the dominant solution of Random Circuits benchmark.}
    \begin{tabular}{|c|c|c|c|c|}
    \hline
        \textbf{Experiment Details} & \textbf{Shots} & \textbf{Mode} & \textbf{M3} & \textbf{Majority Vote} \\ \hline
        Qubits = 20 & 1024 & 3 & 3 & 1 \\ 
        Depth = 318 & 4096 & 3 & 3 & 1 \\ 
        2q Gates = 256 & 16384 & 2 & 1 & 0 \\ \hline
        Qubits = 25 & 2048 & 13 & 7 & 2 \\ 
        Depth = 573 & 6144 & 12 & 7 & 0 \\ 
        2q Gates = 531 & 24576 & 9 & 7 & 2 \\ \hline
        Qubits = 30 & 2048 & 16 & 19 & 11 \\ 
        Depth = 801 & 8192 & 15 & 20 & 18 \\ 
        2q Gates = 848 & 32768 & 18 & 18 & 12 \\ \hline
        Qubits = 40 & 4048 & 18 & 15 & 20 \\ 
        Depth = 1770 & 16384 & 18 & 20 & 21 \\ 
        2q Gates = 2404 & 65536 & 18 & 21 & 20 \\ \hline
    \end{tabular}
    
    \label{tab:RandomCircuits}
\end{table}
\begin{table*}[!htb]
    \centering
    \caption{Dominant bitstrings of unmitigated, M3 mitigated distributions, and QMV using different numbers of shots for the 25-qubit random circuit benchmark.}
    \begin{tabular}{|c|cc|ccc|}
\hline
\multirow{2}{*}{\textbf{Shots}} &
  \multicolumn{2}{c|}{\textbf{Probability of correct bitstring}} &
  \multicolumn{3}{c|}{\textbf{Dominant Bitstrings}} \\ \cline{2-6} 
 &
  \multicolumn{1}{c|}{\textbf{Mode}} &
  \textbf{M3} &
  \multicolumn{1}{c|}{\textbf{Mode}} &
  \multicolumn{1}{c|}{\textbf{M3}} &
  \textbf{Majority Vote} \\ \hline
2048 &
  \multicolumn{1}{c|}{0} &
  0 &
  \multicolumn{1}{c|}{0110010011001001001011011} &
  \multicolumn{1}{c|}{1110001000100000101000001} &
  1010101010101001101010101 \\ \hline
6144 &
  \multicolumn{1}{c|}{0} &
  0 &
  \multicolumn{1}{c|}{0100101011011000110000011} &
  \multicolumn{1}{c|}{1110001000100000101000001} &
  1010101010101010101010101 \\ \hline
24576 &
  \multicolumn{1}{c|}{0} &
  0 &
  \multicolumn{1}{c|}{0011000011111011001000101} &
  \multicolumn{1}{c|}{1010101011000010100000000} &
  1010101010001011101010101 \\ \hline
\end{tabular}

    \label{tab:casestudy}
\end{table*}

\begin{table*}[htb!]
    \centering
    \caption{Dominant bitstrings of unmitigated, M3 mitigated distributions, and QMV using various numbers of shots for the 20 qubit random circuit benchmark.}
    \begin{tabular}{|c|cc|ccc|}
\hline
\multirow{2}{*}{\textbf{Shots}} &
  \multicolumn{2}{c|}{\textbf{Probability of correct bitstring}} &
  \multicolumn{3}{c|}{\textbf{Dominant Bitstrings}} \\ \cline{2-6} 
 &
  \multicolumn{1}{c|}{\textbf{Mode}} &
  \textbf{M3} &
  \multicolumn{1}{c|}{\textbf{Mode}} &
  \multicolumn{1}{c|}{\textbf{M3}} &
  \textbf{Majority Vote} \\ \hline
1024 &
  \multicolumn{1}{c|}{0.0009} &
  0.0011 &
  \multicolumn{1}{c|}{01110101010111110101} &
  \multicolumn{1}{c|}{01110101010111110101} &
  01010101010101010111 \\ \hline
4096 &
  \multicolumn{1}{c|}{0.0003} &
  0.0004 &
  \multicolumn{1}{c|}{01010111000101011101} &
  \multicolumn{1}{c|}{01010111000101011101} &
  01010101010101010111 \\ \hline
16384 &
  \multicolumn{1}{c|}{0.0004} &
  0.0007 &
  \multicolumn{1}{c|}{01010101011101010111} &
  \multicolumn{1}{c|}{01000101010101010101} &
  01010101010101010101 \\ \hline
\end{tabular}

    \label{tab:casestudy20}
\end{table*}
Similar trends are evident in the random circuits benchmark, as depicted in Table \ref{tab:RandomCircuits}. However, notably, there are two instances where
QMV successfully identifies the correct solution, while both Mode and M3 fail. Additionally, all three methods exhibit reduced effectiveness, i.e., higher Hamming distances from the correct solution, with higher depth, qubits, and gates. This decline in effectiveness is attributed to increased noise levels. In the 
high noise regime, almost half of the bits are measured incorrectly, and not much information can be recovered. Next, we look into the reasons why QMV
is effective with detailed case studies.

\begin{table}[h!]
\centering
\caption{Qubit-wise probabilities for the correct bit value for the 20 qubit random circuit benchmark.}
\begin{tabular}{|c|c|ccc|}
\hline
\multirow{2}{*}{\textbf{Qubits}} & \multirow{2}{*}{\textbf{Correct Value}} & \multicolumn{3}{c|}{\textbf{Probability of the correct value}}                                                      \\ \cline{3-5} 
                                 &                                         & \multicolumn{1}{c|}{\textbf{s = 1024}} & \multicolumn{1}{c|}{\textbf{s = 4096}} & \textbf{s = 16384} \\ \hline
0  & 0 & \multicolumn{1}{c|}{0.5009} & \multicolumn{1}{c|}{0.6169}  & 0.6208 \\ \hline
1  & 1 & \multicolumn{1}{c|}{0.8359} & \multicolumn{1}{c|}{0.6809}  & 0.7163 \\ \hline
2  & 0 & \multicolumn{1}{c|}{0.665}  & \multicolumn{1}{c|}{0.701}   & 0.6342 \\ \hline
3  & 1 & \multicolumn{1}{c|}{0.6972} & \multicolumn{1}{c|}{0.6202}  & 0.5508 \\ \hline
4  & 0 & \multicolumn{1}{c|}{0.6582} & \multicolumn{1}{c|}{0.6719}  & 0.7344 \\ \hline
5  & 1 & \multicolumn{1}{c|}{0.6015} & \multicolumn{1}{c|}{0.554}   & 0.5339 \\ \hline
6  & 0 & \multicolumn{1}{c|}{0.6142} & \multicolumn{1}{c|}{0.52783} & 0.5586 \\ \hline
7  & 1 & \multicolumn{1}{c|}{0.5927} & \multicolumn{1}{c|}{0.5499}  & 0.5829 \\ \hline
8  & 0 & \multicolumn{1}{c|}{0.6816} & \multicolumn{1}{c|}{0.5499}  & 0.6049 \\ \hline
9  & 1 & \multicolumn{1}{c|}{0.7988} & \multicolumn{1}{c|}{0.5943}  & 0.6901 \\ \hline
10 & 0 & \multicolumn{1}{c|}{0.5498} & \multicolumn{1}{c|}{0.5664}  & 0.5778 \\ \hline
11 & 1 & \multicolumn{1}{c|}{0.6865} & \multicolumn{1}{c|}{0.6531}  & 0.6537 \\ \hline
12 & 0 & \multicolumn{1}{c|}{0.5654} & \multicolumn{1}{c|}{0.67}    & 0.6841 \\ \hline
13 & 1 & \multicolumn{1}{c|}{0.5947} & \multicolumn{1}{c|}{0.6042}  & 0.6124 \\ \hline
14 & 0 & \multicolumn{1}{c|}{0.58}   & \multicolumn{1}{c|}{0.602}   & 0.549  \\ \hline
15 & 1 & \multicolumn{1}{c|}{0.6865} & \multicolumn{1}{c|}{0.7056}  & 0.6968 \\ \hline
16 & 0 & \multicolumn{1}{c|}{0.6416} & \multicolumn{1}{c|}{0.5932}  & 0.5848 \\ \hline
17 & 1 & \multicolumn{1}{c|}{0.8623} & \multicolumn{1}{c|}{0.6704}  & 0.7123 \\ \hline
\textit{18} & \textit{0} & \multicolumn{1}{c|}{\textit{0.4667}} & \multicolumn{1}{c|}{\textit{0.497}}   & \textit{0.5226} \\ \hline
19 & 1 & \multicolumn{1}{c|}{0.7285} & \multicolumn{1}{c|}{0.7254}  & 0.6632 \\ \hline
\end{tabular}

\label{tab:qubitwise-probabilities}
\end{table}

\subsection{\bf Case study of 25-qubit RC}

\begin{table}[t]
\caption{Comparison in Hamming Distance for the 25-qubit random circuit for AMS QMV approaches. }
\resizebox{\columnwidth}{!}{%
\begin{tabular}{|c|c|c|c|c|c|c|}
\hline
\textbf{Experiment Details} & \textbf{\#shots} & \textbf{Mode} & \textbf{M3} & \textbf{QMV} & \textbf{AMS 0.01} & \textbf{AMS 0.05} \\ \hline
\textbf{Qubits = 25}    & 1536  & 10 & 8  & 5 & 5 & 9 \\ \hline
Untranspiled Depth = 17 & 2048  & 11 & 9  & 6 & 5 & 6 \\ \hline
Transpiled Depth = 137             & 6144  & 9  & 8  & 5 & 4 & 5 \\ \hline
\# 2q Gates = 30          & 24576 & 8  & 14 & 5 & 6 & 6 \\ \hline
\end{tabular}%
}
\label{tab:AMSMV}
\end{table}

Consider the 25-qubit random circuit experiment conducted on the \texttt{ibm\_sherbrooke} quantum device, as detailed in Table \ref{tab:casestudy}. In this scenario, the noise-free solution is an alternating sequence of zeros and ones, ``1010101010101010101010101." Our goal is to assess the performance of the M3 and QMV methods within low-shot regimes.

Initially, with 2048 shots, QMV produces 2 incorrect bits, while the mode and M3 methods yield 13 and 7 incorrect bits, respectively. However, as the number of shots increases, QMV manages to identify the correct bitstring, whereas the mode distribution fails to identify the correct solution. In this case, M3 struggles to recover the correct solution as it amplifies incorrect bitstrings.

The reason behind M3's inability to derive the correct bitstring is that M3 relies on calibration data to derive transition probabilities among observed bitstrings. However, in a low shot regime, the correct bitstring may never be observed. Moreoever, even if the correct bitstring
is observed, then it is likely observed only 1--2 times. 

If the correct bitstring is never observed, then M3 cannot obtain the correct solution, because the matrix reduction step in M3 only considers observed bitstrings. In our case, as the correct solution was not observed, both mode and M3 were unable to identify it. Hence, QMV demonstrates its ability to identify the correct solution even when the correct bitstring is never observed. Next, let us see what happens when the correct bitstring is observed but is not the dominant one.

\subsection{\bf Case study of 20-qubit RC}

Consider a case study with 20 qubits, where the correct bitstring is observed but does not dominate. The details for this case study are available in Table \ref{tab:casestudy20}. Theoretically, M3 should identify the correct bitstring as long as it is observed. However, the limited number of shots hampers M3's accuracy in correcting the bitstring. This limitation arises from the insufficient transitions captured in the reduced transition matrix,
owing to the limited number of shots. The insufficient transitions compel the M3 algorithm to bolster the most dominant bitstring, which unfortunately is not the correct one.

In our example, with 1024 and 4096 shots, M3 primarily increases the probability of the most dominant bitstring. 
At higher shot counts, M3 starts enhancing bitstrings closer in Hamming distance to the correct one rather than only boosting the mode. 
With the shot count reaching 16384, M3 amplifies the correct solution probability by 75\%. However, despite M3 beginning to enhance 
bitstrings closer in Hamming distance to the correct solution, the correct one was not boosted sufficiently, and M3 remains outclassed by QMV in this case study. Not only does QMV identify the solution with the lowest Hamming distance among all methods, but it also identifies the correct result when using 16384 shots. Another interesting observation is that, generally, the same sets of qubits remain incorrect for different numbers of shots. As the number of shots increases, the
qubits tend to be gradually corrected by QMV. This, however, is not the case for M3 and mode. Consequently, we posit that we can determine which qubits are 
likely wrong by looking at QMV's data.

The qubit-wise probabilities depicted in Table \ref{tab:qubitwise-probabilities} shed light on why QMV is successful. This table represents probabilities for the 20-qubit RC benchmark in this case study. Qubit 18 (highlighted in italic fonts) suffers from an incorrect vote when the number of shots is 1024 or 4096. When the number of shots increases to 16384, QMV provides the correct solution, but the vote between 0 and 1 remains close. This suggests that this qubit likely suffers from high measurement errors.   
Under such circumstances we suggest using measurement subsetting.
\subsection{\bf Adaptive measurement subsetting results}

We follow the AMS approach discussed in Section IV.D on the 25-qubit random circuit benchmark executed on the quantum device, \texttt{ibm\_sherbrooke}. Table \ref{tab:AMSMV} presents the results.  The thresholds used to select the qubits with close voting were set to 0.01 and 0.05, and the detailed information on which qubits are selected for measurement subsetting is shown in Table \ref{tab:AMSCloseVote}. The results indicate that AMS can reduce the Hamming distance to the noise-free solution while using the same number of shots as QMV. However, the choice of threshold proves crucial. Lower thresholds let fewer qubits benefit from measurement subsetting. Opting for a higher threshold, seems to worsen the Hamming distance, as seen in the AMS 0.05 column of Table \ref{tab:AMSMV}. 
The reason for this worsened Hamming distance is that a high threshold leads to a high number of qubits undergoing measurement subsetting. This, in turn, decreases the number of shots per circuit, making each qubit more susceptible to noise. 
As shown in Table \ref{tab:AMSCloseVote}, when the number of shots is 1536, 768 are designated for the full circuit measurement run, resulting in 12 qubits being considered close vote qubits for AMS 0.05, with only 1 qubit identified as a close vote qubit for AMS 0.01. Consequently, AMS 0.05 has only 64 shots for each measurement subsetting circuit, leading to a deterioration in the performance of QMV. To avoid such scenarios, a simple rule of thumb is to select a threshold such that each measurement subsetting circuit has more than 100 shots. As can be seen from Table \ref{tab:AMSMV} and Table \ref{tab:AMSCloseVote}, with this rule, AMS improves the Hamming distance compared to QMV. But the improvement is limited, because AMS only mitigates measurement errors,
while our benchmark suffers more from gate errors (137 gates) than from measurement errors.

\begin{table}[!htb]
\centering
\caption{Close vote qubits and shots per circuit for 25 qubit random circuit when using Adaptive Measurement Subsetting.}
\begin{tabular}{|c|LL|cc|}
\hline
\multirow{2}{*}{\textbf{\# shots}} & \multicolumn{2}{c|}{\textbf{Close vote qubits}} & \multicolumn{2}{c|}{\textbf{Shots per circuit}} \\ \cline{2-5} 
      & \multicolumn{1}{c|}{\textbf{AMS 0.01}} & \textbf{AMS 0.05}                                & \multicolumn{1}{c|}{\textbf{AMS 0.01}} & \textbf{AMS 0.05} \\ \hline
1536  & \multicolumn{1}{c|}{{[}8{]}}          & {[}22, 20, 17, 14, 13, 12, 10, 9, 8, 4, 1, 0{]} & \multicolumn{1}{c|}{768}              & 64               \\ \hline
2048  & \multicolumn{1}{c|}{{[}22, 17, 8{]}}  & {[}22, 20, 17, 14, 13, 9, 8, 1, 0{]}            & \multicolumn{1}{c|}{341}              & 113              \\ \hline
6144  & \multicolumn{1}{c|}{{[}22, 17{]}}     & {[}22, 20, 17, 14, 9, 8, 4, 1, 0{]}             & \multicolumn{1}{c|}{1536}             & 341              \\ \hline
24576 & \multicolumn{1}{c|}{{[}22, 17{]}}     & {[}22, 20, 17, 14, 12, 9, 8, 1, 0{]}            & \multicolumn{1}{c|}{6144}             & 1365             \\ \hline
\end{tabular}
\label{tab:AMSCloseVote}
\end{table}

\section{Conclusion} \label{sec:conculsion}

In this paper, we show that a straightforward {\em qubit-wise majority vote} (QMV) scheme
identifies the correct bitstring for quantum algorithms with a single correct output.
We focus on the low-to-medium shot regime on a real superconducting quantum device. 
Our evidence highlights the capability of QMV to identify the correct output, even when it is never observed. 
This contrasts with methods such as M3, which rely on observing the correct output, and fail when it is never observed. 
Furthermore, QMV can also indicate that some qubits are likely to be more erroneous than others. 

We saw in Section~\ref{sec:main} that QMV requires a number of shots logarithmic in the number of qubits, $n$,
which greatly improves over the exponential requirements of M3. The mode requires even more shots than M3.
However, our numerical results on \texttt{ibm\_sherbrooke}, an IBM quantum device, were less conclusive.
We believe that the less conclusive nature of our numerical results is due to some of the qubits being noisier
than others. In contrast, our analysis for the number of shots required by QMV (Section~\ref{subsec:num_shots_derivation})
assumes i.i.d. noise. Indeed, it seems important to focus future work on error mitigation methods that are robust to
a wide range of noise models.
Moreover, because QMV is a classical post-processing step that is run on quantum data, it may be possible to
apply QMV to classical applications, and apply other related classical schemes to quantum error mitigation.
Finally, QMV demonstrates that having prior knowledge about the nature of the output of a quantum algorithm can 
lead to simple and effective error mitigation schemes.

\section*{Acknowledgment}

The authors thank Wladimir Silva for helpful discussions.
This work is partly funded by NSF grants 1818914 and 2325080
(with a subcontract to NC State University from Duke University), and 2120757 (with a subcontract to NC State University
from the University of Maryland).

\bibliographystyle{IEEEtranS}
\bibliography{refs}

\vspace{12pt}

\end{document}